\documentclass[dvipdfmx,12pt]{article}
\usepackage{amsmath,amsfonts,amssymb,booktabs,multirow,cite}
\usepackage{array}
\usepackage[dvipdfmx]{graphicx}
\usepackage{xcolor}
\usepackage{float}
\usepackage[unicode]{hyperref}
\hypersetup{
colorlinks=false,
linkbordercolor=blue,
citebordercolor=red,
urlbordercolor=cyan
}

\newcommand{\del}{\partial}%

\newcommand{\half}{\frac{1}{2}}%

\newcommand{\nn}{\nonumber}

\makeatletter
\@addtoreset{equation}{section}
\renewcommand{\theequation}{\thesection.\@arabic\c@equation}
\makeatother

\begin{document}


\begin{titlepage}

\vspace*{-15mm}   
\baselineskip 10pt   
\begin{flushright}   
\begin{tabular}{r}    
\end{tabular}   
\end{flushright}   
\baselineskip 24pt   
\vglue 10mm   

\begin{center}
{\Large\bf
 {Quantum focusing conjecture in two-dimensional evaporating black holes}
}

\vspace{8mm}   

\baselineskip 18pt   

\renewcommand{\thefootnote}{\fnsymbol{footnote}}

Akihiro Ishibashi${}^{1,2}$\footnote[2]{akihiro@phys.kindai.ac.jp}, Yoshinori Matsuo${}^{1}$\footnote[3]{ymatsuo@phys.kindai.ac.jp} and Akane Tanaka${}^{1}$\footnote[4]{2333310152s@kindai.ac.jp} 

\renewcommand{\thefootnote}{\arabic{footnote}}
 
\vspace{5mm}   

{\it  
${}^{1}$Department of Physics and ${}^{2}$Research Institute for Science and Technology, \\ 
Kindai University, Higashi-Osaka, Osaka 577-8502, Japan
}
  
\vspace{10mm}   

\end{center}

\begin{abstract}
We consider the quantum focusing conjecture (QFC) for two-dimensional evaporating black holes in the Russo-Susskind-Thorlacius (RST) model. The QFC is closely related to the behavior of the generalized entropy.
In the context of the black hole evaporation, the entanglement entropy of the Hawking radiation is decreasing after the Page time, and therefore it is not obvious whether the QFC holds.
One of the present authors previously addressed this problem in a four-dimensional spherically symmetric dynamical black hole model and showed that the QFC is satisfied. However, the background spacetime considered was approximated by the Vaidya metric, and quantum effects of matters in the semiclassical regime were not fully taken into consideration. It remains to be seen if the QFC in fact holds for exact solutions of the semiclassical Einstein equations. In this paper, we address this problem in the RST model, which allows us to solve the semiclassical equations of motion exactly. 
We prove that 
the QFC is satisfied for evaporating black holes in the RST model with the island formation taken into account. 
\end{abstract}

\end{titlepage}



\tableofcontents


\section{Introduction}
In general relativity, the focusing theorem is key to understanding the basic properties of gravitation. By combining the Raychaudhuri equation and certain energy conditions, the focusing theorem plays a central role in establishing various important results in general relativity, such as the singularity theorems~\cite{Penrose:1964wq}. For example, the black hole second law or the area theorem is based on the null focusing theorem, which asserts that under the null energy condition (NEC), the expansion $\theta$ of a null geodesic congruence is non-increasing:   
\begin{align}
\label{nullfocus}
\frac{d\theta}{d\lambda} \leq 0 \,,  
\end{align}  
where $\lambda$ is an affine parameter of a null geodesic, say $\gamma$, and  
\begin{align}
\label{def:class:exp}
\theta:= \dfrac{1}{{\cal A}}\dfrac{d {\cal A}}{d\lambda } \,, 
\end{align}
with $\cal{A}$ being the sectional area of the null congruence including $\gamma$. Then, applying this theorem to a black hole with $\cal A$ regarded as the horizon area and using global techniques in general relativity, one can show that
\begin{align} 
\label{arealaw}
d{\cal A} \geq 0 \,. 
\end{align} 
However, when quantum effects are considered, the null focusing theorem does not in general hold due to the violation of 
the NEC\footnote{ 
Any locally defined energy conditions could, in general, be violated by quantum field effects. As an alternative to such a locally defined energy condition, the averaged null energy conditions (ANEC) was proposed, and the focusing theorem was reformulated under the ANEC [see e.g., \cite{Borde:1987qr}]. There have been extensive studies on the ANEC [see, e.g., Refs~\cite{Klinkhammer:1991ki,Wald:1991xn,Ford:1994bj,Graham:2007va,Flanagan:1996gw,Kelly:2014mra,Faulkner:2016mzt,Hartman:2016lgu,Urban:2009yt,Visser:1994jb,Ishibashi:2019nby,Iizuka:2019ezn,Rosso:2020cub,Iizuka:2020wuj,Ishibashi:2021hqx} and references therein].
}. 

As an alternative notion which is applicable even to the semiclassical regime, the quantum focusing conjecture (QFC) has been proposed~\cite{Bousso:2015mna}. 
The basic idea of the QFC is motivated from Bekenstein's generalized entropy $S_{\rm gen}$, defined as~\cite{Bekenstein:1972tm,Bekenstein:1973ur}, 
\begin{align}
S_{\text{gen}} = \frac{{\cal A}}{4G_{N}} + S_{\text{out}} \,,
\label{generalizedentropyorigine}
\end{align}
where $S_{\text{out}}$ is the von Neumann entropy of matter fields or radiation outside the black hole. 
As a semiclassical generalization of the classical area law \eqref{arealaw}, the generalized second law (GSL) asserts that $S_\text{gen}$ is non-decreasing:
\begin{align}
dS_{\text{gen}} \geq 0 \,. 
\end{align}
Then, by using the generalized entropy, the null focusing theorem~\eqref{nullfocus} is refined to the QFC~\cite{Bousso:2015mna}, which states: 
\begin{align}
\label{def:QFC}
\frac{d\Theta}{d\lambda} \leq 0 \,, 
\end{align}
with $\Theta$---called the quantum expansion---being defined as a quantum generalization of the expansion~\eqref{def:class:exp}. 
For the present context, $\Theta$ is given by 
\begin{align}
\label{def:Qexp}
\Theta = \frac{4G_{N}}{{\cal A}} \frac{dS_{\text{gen}}}{d\lambda} \,. 
\end{align} 

The formula~\eqref{def:QFC} with \eqref{def:Qexp} applies not only for a black hole horizon as discussed above but also for any co-dimension two surface with the area ${\cal A}$ on a Cauchy hypersurface, for which $S_{\rm out}$ in \eqref{generalizedentropyorigine} is given as the entanglement entropy for quantum fields on one side of the Cauchy hypersurface divided by the surface.

The QFC has been shown to hold in various situations in which quantum field effects violate the classical null focusing theorem. It is expected that the QFC can be used to establish many important results in semiclassical quantum gravity, just like the classical focusing theorem is in general relativity. For instance, the QFC is used to derive the quantum null energy condition (QNEC)~\cite{Bousso:2015wca,Koeller:2015qmn,Fu:2017evt}, which is a quantum generalization of the NEC. 
An improved version of the Bousso bound 
\cite{Bousso:1999xy,Flanagan:1999jp,Strominger:2003br,Bousso:2003kb,Bousso:2014sda,Bousso:2014uxa} 
also follows from the QFC.%
\footnote{%
The Bousso bound in the two-dimensional dilaton gravity was also studied in 
\cite{Strominger:2003br,Franken:2023ugu}. 
}

One of the most important problems in quantum gravity is the black hole information paradox. The black hole evaporation due to Hawking radiation implies that an initial pure quantum state forming a black hole appears to evolve into a mixed state. This picture, however, contradicts the unitary evolution in quantum theory, provided that the Hawking radiation is perfectly thermal and the black hole evaporates completely. A key for resolving this paradox is the Page curve~\cite{Page:1993wv,Page:2013dx}, which describes the behavior of the entanglement entropy of the Hawking radiation under quantum unitary evolution: the entanglement entropy initially monotonically increasing should turn to decrease at some point---called the Page time---and eventually vanishes at the end of the evaporation. One of the recent advances along this line is the proposal of the so called ``island''~\cite{Penington:2019npb,Almheiri:2019psf,Almheiri:2019hni,Almheiri:2019yqk,Penington:2019kki,Almheiri:2019qdq} which is a certain region supposed to host part of the degrees of freedom of the Hawking radiation in late time, despite being typically located inside the evaporating black hole. By using the idea of islands, a number of studies have been done \cite{Chen:2019uhq,Almheiri:2019psy,Akers:2019lzs,Marolf:2020xie,Balasubramanian:2020hfs,Gautason:2020tmk,Anegawa:2020ezn,Hashimoto:2020cas,Sully:2020pza,Hartman:2020swn,Hollowood:2020cou,Krishnan:2020oun,Alishahiha:2020qza,Chen:2020uac,Almheiri:2020cfm,Geng:2020qvw,Bousso:2020kmy,Krishnan:2020fer,Chen:2020tes,Hartman:2020khs,Balasubramanian:2020xqf,Chen:2020hmv,Hernandez:2020nem,Geng:2020fxl,Geng:2021wcq,Geng:2021hlu,Bousso:2021sji}.

It is of considerable interest in studying possible roles of the QFC in the context of the black hole information paradox, as the QFC closely connects classical geometry and quantum field effects. However, according to the above observation on a Page curve and an island formation, both the entanglement entropy of the Hawking radiation and the area of the evaporating black hole horizon decrease after the Page time. This begs the question of whether or not the QFC holds in the course of black hole evaporation, in particular, after the Page time.  

In order to incorporate quantum field effects, 
we calculate the generalized entropy by using the formula of islands. 
We consider the QFC for a congruence of outgoing null geodesics outside the horizon. 
At early times, the QFC can be studied by using the ordinary definition of the generalized entropy.
After the Page time, we need to take islands into calculations to reproduce 
the non-trivial behavior of the entanglement entropy. 
Note that our analysis should be distinguished from 
applications of the QFC for determining locations of islands 
\cite{Almheiri:2019yqk,Akers:2019lzs,Hartman:2020khs,Bousso:2021sji}. 
The position of islands should be chosen so that the entanglement entropy is extremized, 
and hence, can be seen by calculating the quantum expansion. 
In such studies, a candidate of a boundary of islands moves along a null geodesic congruence, 
and the QFC for it is considered. 
In contrast, we study the congruence in outer places, and islands appear 
in another place than the congruence. 
We consider the time evolution of the generalized entropy along the null geodesic congruence, 
and the position of islands is determined so that the generalized entropy 
is extremized in each moment in the time evolution. 

One of the present authors previously considered this issue in a four-dimensional spherically symmetric model~\cite{Matsuo:2023cmb}. It is in general very hard to setup the background geometry describing a black hole evaporation by solving the semiclassical Einstein equations. First of all, the vacuum expectation value $\langle T_{\mu \nu} \rangle$ of stress-energy tensor for an arbitrary background spacetime is necessary to construct the semiclassical Einstein equations. 
It is practically difficult to calculate $\langle T_{\mu \nu} \rangle$ for quantum fields in four-dimension. Second, even when an expression of $\langle T_{\mu \nu} \rangle$ is obtained, solving the semiclassical Einstein equations is still a difficult task due to the fact that $\langle T_{\mu \nu} \rangle$ involves, in general, higher order derivatives. For these reasons, the analysis of~\cite{Matsuo:2023cmb} has been done by making several assumptions to sufficiently simplify the problem. In particular, the background geometry is not a solution of the semiclassical Einstein equations but quantum effects of matter are only partially taken into account: the Vaidya metric was exploited to include the negative incoming energy of quantum vacuum state, on which the QFC was shown to be satisfied. However, it is not clear whether such a background model can be justified in the semiclassical context. 

In order to critically examine the QFC, it is most desirable to consistently solve the semiclassical Einstein equations. This is a formidable task as explained above, but can be undertaken---in fact, analytically---in two dimensions. It is well known that two dimensional gravity with dilaton fields~\cite{Callan:1992rs,Russo:1992ax,Gautason:2020tmk} admits non-trivial dynamical black hole solutions, which enjoy many of the features of black holes in four-dimensions, such as the uniqueness and thermodynamic analogy~\cite{Frolov:1992xx}. 
In this paper, we consider the two-dimensional Russo-Susskind-Thorlacius (RST) dilaton gravity~\cite{Russo:1992ax}, whose semiclassical Einstein equation is solvable analytically, and show that the QFC indeed holds for dynamical black holes in the entire course of the Hawking evaporation with the island formation taken into consideration.

This paper is organized as follows. In the next section, we briefly review the black hole solutions in the two-dimensional gravity and generalized entropy with and without an island based on~\cite{Gautason:2020tmk}. Then, in section~\ref{qfc}, we prove that the quantum focusing conjecture holds in two dimensional evaporating black hole. In section~\ref{conclusion}, we present conclusion and outlook.

\section{2D black holes and Islands}\label{2D}
In this section, we briefly review two-dimensional black holes in the RST model 
\cite{Callan:1992rs,Russo:1992ax,Gautason:2020tmk} 
and the island rule to calculate the generalized entropy of Hawking radiation in the RST model.

\subsection{RST model}
We first consider the CGHS model\cite{Callan:1992rs}, 
in which black holes are studied in the two-dimensional dilaton gravity with the classical action given by
\cite{Witten:1991yr,Elitzur:1990ubs,Mandal:1991tz,Dijkgraaf:1991ba,Callan:1992rs} 
\begin{align} 
\label{def:action:CGHS} 
I_{\text{CGHS}} = \frac{1}{2\pi} \int d^2x \sqrt{-g}e^{-2\phi}\{R + 4(\nabla \phi)^2 + 4\lambda^2\} \,,
\end{align}
where 
$R$ is the Ricci scalar, 
$\phi$ is a dilaton field and $\lambda$ is a parameter characterizing the length scale.
We can set $\lambda=1$ by an appropriate rescaling of two-dimensional coordinates. 
We also introduce matter fields. 
For simplicity, we focus on conformal matters, and then, the energy-momentum tensor 
has the conformal anomaly for curved background when quantum effects are taken into consideration: 
\begin{align}
\langle T^{\mu}{}_{\mu} \rangle = \frac{c}{12}R \,, 
\label{semi energy-momentum trace}
\end{align}
where 
$c$ is the central charge.
In the conformal gauge, the metric takes the following form: 
\begin{align}
ds^2 = - e^{2\rho}dx^{+}dx^{-}\,,
\end{align}
where $\rho$ is a function of $x^{+}$ and $x^{-}$. 
In this gauge, 
eq.~\eqref{semi energy-momentum trace} is expressed as 
\begin{equation}
 T_{+-}
 = -\frac{c}{6}\del_{+}\del_{-}\rho \,. 
 \label{T+-}
\end{equation}
The other components can be calculated by integrating the conservation law $\nabla^\mu T_{\mu\nu}=0$ as 
\begin{align}
T_{\pm\pm} = 
\frac{c}{12}\left[2\del^2_{\pm}\rho - 2(\del_{\pm}\rho)^2 + t_{\pm}\right] \,, 
\label{energy-moumentum com}
\end{align}
where $t_{\pm}(x_{\pm})$ are integration constants determined by physical boundary conditions. 
%
%
It should be noted that $t_\pm$ transforms under the coordinate transformation as 
\begin{equation}
 \left(\frac{\sigma^\pm}{w^\pm}\right)^2 t_\pm(\sigma^\pm) 
 = 
 t_\pm(w^\pm) + \{\sigma^\pm,w^\pm\} \ , 
 \label{t-trans}
\end{equation}
where 
\begin{equation}
 \{f(x),x\} 
 = 
 \frac{f'''}{f'} - \frac{3}{2} \frac{\left(f''\right)^2}{\left(f'\right)^2} 
\end{equation}
is the Schwarzian derivative. 


The quantum energy-momentum tensor \eqref{T+-}--\eqref{energy-moumentum com} 
can be reproduced by adding 
the following non-local Polyakov term to the action $I_{\text{CGHS}}$, \eqref{def:action:CGHS}: 
\begin{align}
I_{Q}= -\frac{c}{12\pi} \int dx^{+}dx^{-} \del_{+}\rho \del_{-}\rho \,. 
\end{align}
However, this term 
breaks a symmetry of the classical action \eqref{def:action:CGHS}: 
\begin{equation}
 \delta\phi = \delta\rho = \varepsilon e^{2\phi} \,. 
\label{sym}
\end{equation}
In order to preserve this symmetry, 
we further modify the action by adding the following term~\cite{Russo:1992ax},
\begin{align}
I_{\text{RST}} = - \frac{c}{48\pi} \int d^2x \sqrt{-g} \phi R \,. 
\label{IRST}
\end{align}
By using the symmetry \eqref{sym}, we can set $\rho=\phi$ without loss of generality. 
The null coordinates $x^\pm$ in this gauge correspond to the Kruskal coordinates. 
Following \cite{Russo:1992ax,Bilal:1992kv}, we introduce a new field variable: 
\begin{align}
\label{def:Omega}
\Omega = e^{-2\phi} + \frac{c}{24}\phi \,, 
\end{align}
which significantly simplifies the field equations. The field variable $\Omega$ takes a lower bound at $\Omega=\Omega_{\text{crit}} = ({c}/{48})[1- \log({c}/{48})]$, which corresponds to the boundaries of spacetime. The field equations for the action $I_{\text{CGHS}}+I_{\text{Q}} + I_{\text{RST}}$ can be expressed as 
\begin{align}
\del_{+}\del_{-}\Omega + 1 = 0 \,, \qquad -\del_{\pm}^2\Omega = \frac{c}{24}t_{\pm} \,, 
\label{field equation RST}
\end{align}
where $t_{\pm}$ is the same 
integration constants 
as before.

Note that $t_\pm$ is not the energy-momentum tensor. 
The off-diagonal components of the energy-momentum tensor is given by \eqref{energy-moumentum com} 
and the equation of motion can be expressed in terms of $\rho$ and $\phi$ as 
\begin{equation}
 \left(e^{-2\phi} - \frac{c}{48}\right)
 \left[ \partial_\pm^2 \phi - 2 \partial_\pm \rho \partial_\pm\phi \right] 
 = - \frac{1}{2} T_{\pm\pm} \ . 
 \label{QEM}
\end{equation}

\subsubsection{Linear dilaton vacuum}
The linear dilaton vacuum solution takes the form
\begin{align}
\Omega = -x^{+}x^{-} - \frac{c}{48}\log(-x^{+}x^{-}) \,,
\label{ld-Omega}
\end{align}
where we have assumed $T_{\pm\pm}=0$. Substituting the solution $\Omega$ into the second equation of \eqref{field equation RST}, 
we find that 
\begin{align}
t_{\pm} = -\half \frac{1}{(x^{\pm})^2} \,.
\label{t-ld}
\end{align}
From \eqref{def:Omega}, the dilaton $\phi$ can be read off as 
\begin{equation}
 e^{-2\phi} = - x^+ x^- \,. 
\end{equation}
Since we have taken the gauge condition $\rho=\phi$ the metric is expressed as 
\begin{align}
ds^2 = \frac{1}{x^{+}x^{-}} dx^{+}dx^{-} \,, 
\label{dilatonvac}
\end{align}
%
which is nothing but flat spacetime. 
In order to see this more explicitly, we introduce the coordinates $(\sigma^+,\sigma^-)$ 
which are defined by $x^+ = e^{\sigma^+}$ and $x^- =- e^{\sigma^-}$. 
Then, the metric takes the standard form of flat spacetime in double null coordinates: 
\begin{align}
 ds^2 = -d\sigma^{+} d\sigma^{-} \,.
\end{align}
The energy-momentum tensor is determined by the equation of motion \eqref{QEM} 
and gives $T_{++}=T_{--}=0$. 
Thus, this solution is a vacuum solution. 
Since the dilaton is proportional to the spatial coordinate $\sigma = \frac{1}{2}(\sigma^+-\sigma^-)$: 
\begin{equation}
 \phi = - \sigma \,, 
\end{equation}
this solution is called a linear dilaton vacuum solution.

\subsubsection{Two-sided eternal black hole}
The eternal black hole solution is given by 
\begin{align}
\Omega= M(1-uv) + \Omega_{\text{crit}} \,, 
\end{align}
where the coordinate system was set to $x^{+}=\sqrt{M}v , \; x^{-}=\sqrt{M}u$. 
The parameter $M$ is an integration constant and proportional to the black hole mass. 
In order to see that this solution describes the eternal black hole solution, we consider the classical limit. 
From \eqref{def:Omega}, the dilaton can be read off as 
\begin{align}
e^{-2\phi} = M(1-uv) + \mathcal O(\epsilon) \ ,  
\end{align} 
where we have introduced the parameter $\epsilon := c/48M$. 
We assume $M\gg c$, or equivalently $\epsilon\ll 1$ in the rest of the paper, 
implying that quantum effects are much smaller than the classical mass of the black hole. 

The metric is approximately expressed as 
\begin{align}
ds^2 = -\frac{dvdu}{1-vu} + \mathcal O(\epsilon) \,. 
\label{bh-metric}
\end{align}
A curvature singularity is located at $uv=1$, and the event horizon is at $uv=0$. 
It is straightforward to see that this black hole solution is asymptotically flat 
by using the coordinates $(\sigma^+,\sigma^-)$. 
The global structure of this solution is essentially the same as that of the two-dimensional part of the Schwarzschild black hole. 

We can calculate the energy-momentum tensor by using \eqref{QEM} 
and find $T_{++}=T_{--}=0$ at the leading order of the small $\epsilon$ expansion, 
or equivalently in the classical limit. 
Quantum effects in the energy-momentum tensor can be obtained by considering 
the next-to-leading order terms. 
Alternatively, we can estimate quantum corrections from \eqref{energy-moumentum com}. 
We find $t_{\pm}(x^{\pm})=0$ from \eqref{field equation RST}. 
The parameter $t_\pm$ transforms under the coordinate transformation as \eqref{t-trans}. 
If the coordinate system is transformed to $(\sigma^+,\sigma^-)$-coordinates, 
we obtain $t_{\pm}(\sigma^{\pm}) = 1/2$, 
which can be expressed as $T_{\sigma^\pm \sigma^\pm}(\sigma^{\pm})={c}/{24}$ 
near the spatial infinity since $\rho\simeq 0$ and $T_{\sigma^\pm \sigma^\pm} \simeq c t_\pm(\sigma^\pm)/12$ there. 
This implies that the black hole has the temperature $T= {1}/{2\pi}$, which is independent of the mass.

\subsubsection{Dynamical black hole}
We consider a situation where a shock wave of 
matter fields
is injected into a linear dilaton vacuum. Here, we assume 
that the classical part of the energy-momentum tensor is given by 
\begin{align}
 T_{++}&= 
 \frac{2M}{x_0^+}
 \delta(x^{+} - x^{+}_{0}) \,, &
 T_{--}&=0 \,.
\end{align}
If we assume that the spacetime is flat in the infinite past, then $t_\pm$ is given by \eqref{t-ld}. 
before the injection of the shock wave at $x_0^+$. 
The shock wave introduces an additional delta-functional term to $t_+$, 
and $t_\pm$ becomes
\begin{align}
 t_+ 
 &= 
 - \frac{1}{2} \dfrac{1}{(x^+)^2} + \frac{24 M}{c\, x_0^+} \delta (x^+-x_0^+) \,,
&
 t_- 
 &= 
 - \frac{1}{2} \dfrac{1}{(x^-)^2} \,.
\end{align}
Then, 
the solution $\Omega$ is expressed as  
\begin{align}
\Omega &= -x^{+}x^{-} - \frac{c}{48}\log(-x^{+}x^{-}) 
- \frac{M}{x_0^+}
(x^{+}-x^{+}_{0})\theta(x^{+}-x^{+}_{0}).
\end{align}
We define the coordinates $(u,v)$ as 
\begin{align}
\label{coord:xpm}
x^{+}= x^{+}_{0}v \,, \qquad x^{-}=\frac{M}{x^{+}_{0}}u \,,
\end{align}
and then, 
$\Omega$ is rewritten for $v>1$ as 
\begin{align}
\Omega= M\left[1-v(u+1)-\epsilon\log(-Muv) \right] \,,
\label{v>1}
\end{align}
and for $v<1$ as 
\begin{align}
\Omega=M\left[-vu-\epsilon\log(-Muv) \right]. 
\label{v<1}
\end{align}
The solution \eqref{v<1} for $v<1$ is nothing but the linear dilaton solution \eqref{ld-Omega}, 
and the spacetime is flat. 
The solution \eqref{v>1} is a black hole solution. 
In the small $\epsilon$ expansion, the metric for \eqref{v>1} is expressed as 
\begin{align}
ds^2 
&= \frac{1}{1-v(u+1)}dudv + \mathcal O(\epsilon) \,.
\label{dynamicalbh}
\end{align}
By introducing coordinates $\tilde u= u+1$, the metric\eqref{dynamicalbh} 
is further rewritten as 
\begin{equation}
 ds^2 = \frac{1}{1-v\tilde u}d\tilde u dv + \mathcal O(\epsilon) \,.
\end{equation}
This has exactly the same form to the eternal black hole metric \eqref{bh-metric}. 
As in the case of the eternal black hole solution, the asymptotically flat structure 
can be seen by introducing the coordinates 
\begin{align}
 v &= e^{\tilde{\sigma}^+} \,,
 &
 \tilde u &= - e^{\tilde{\sigma}^-} \,.
\end{align}
Note that $\tilde{\sigma}^+$ is the same as $\sigma^+$ but 
$\tilde{\sigma}^-$ is different from $\sigma^-$ since 
$u$ coordinate is shifted to $\tilde u$ coordinate at the injection of the shock wave. 
Thus, we have obtained a classical solution that describes an evolution from the linear dilaton vacuum \eqref{dilatonvac} to the black hole spacetime \eqref{dynamicalbh}. 
Hereafter we are mainly interested in the black hole region $v>1\ (x^+>x_0^+)$.

%

\subsection{Generalized entropy}
In order to study the QFC, we calculate the generalized entropy \eqref{generalizedentropyorigine}. 
We consider a point $A$ which divide a Cauchy surface to two connected regions. 
The black hole, or equivalently the event horizon is located in one of the two regions, 
and we calculate the generalized entropy on the other region. 
We refer to the point $A$ as the anchor point and call the region 
with (without) the black hole the interior (exterior) of $A$. 
We assume that the quantum state on the Cauchy surface is pure, 
and von Neumann entropy of the exterior is the entanglement entropy. 
The island rule states that the entanglement entropy of the exterior of $A$, 
including the gravity part, is given by 
\begin{equation}
 S_\text{gen} 
 = 
 \mathrm{min}
 \left\{
 \mathrm{ext}
 \left[
 \sum_{A,I}\frac{\text{Area}}{4G_N} 
 + S_\text{bulk}
 \right]
 \right\}
 \,,
\end{equation}
where $S_\text{bulk}$ is the entanglement entropy on the union 
of the exterior of $A$ and an additional region which is called islands. 
The area terms are the sum of all area of $A$ and $I$, 
where $I$ stands for boundaries of islands. 
We should consider all possible configurations of islands 
which extremize the entanglement entropy. 
Islands can consist of multiple connected regions or be empty (no-island), 
but only two cases of a single connected region and empty have extrema of the entanglement entropy. 
These two cases will be referred to as the island configuration and no-island configuration, respectively. 

The generalized entropy in the island configuration 
is expressed as \cite{Gautason:2020tmk}  
\begin{align}
\label{def:S_gen}
S_{\text{gen}} = 
\frac{\text{Area(A)}}{4G_{N}} + 
\frac{\text{Area(I)}}{4G_{N}} + S_{\text{bulk}}[\mathcal{S}_{\text{AI}}] \,. 
\end{align}
The first term on the right hand side 
corresponds to the area term in the higher dimensional cases, 
which is identified as 
\begin{equation}
 \frac{\text{Area}}{4G_{N}} = 2(\Omega-\Omega_{\text{crit}})  
 \label{area}
\end{equation}
for the RST model.%
\footnote{%
There are two different interpretations of the area term of the entanglement entropy. 
One is simply to identify $\Omega$ with the area term \cite{Gautason:2020tmk}, 
which is consistent with the Wald entropy for the RST model. 
The other is to identify $\Omega$ as the total entropy including the matter part 
\cite{Fiola:1994ir,Strominger:2003br,Hartman:2020swn}. 
The RST action contains the effective action for the Weyl anomaly, and 
the area term is obtained by subtracting contributions from the Weyl anomaly in the second definition. 
In this paper, we adopt the first definition. 
} 
In fact, $\Omega$ is the Noether charge or Wald entropy~\cite{Wald:1993nt} 
derived from the action $I_{\text{CGHS}}+I_Q+I_{\text{RST}}$ as shown in~\cite{Myers:1994sg}.
This term was ignored in the study on the Page curve \cite{Gautason:2020tmk}, 
but plays important role in the QFC. 
We will omit this term in this section, but take it into account explicitly in the next section. 
The second term is the area term at the boundary of the island, which is 
$2(\Omega(I)-\Omega_{\text{crit}})$, arises from effects of islands and 
vanishes if there is no island. The third term, $S_{\text{bulk}}$, is the von Neumann entropy of the CFT matter fields
on 
(the complement of) 
a spacelike surface $\mathcal{S}_{AI}$, connecting the boundary of island $I$ and the anchor point $A$.
The bulk entropy $S_{\text{bulk}}$ is expressed as
\begin{align}
S_{\text{bulk}}[\mathcal S_{\text{AI}}] = \frac{c}{6}\log\left|d(A,I)^2 e^{\rho \text{(A)}}e^{\rho\text{(I)}}\right|_{t_\pm=0} \,,
\end{align}
where $d(A,I)$ is the distance between the anchor curve and the island, 
\begin{equation}
 d(A,I)^2 = \left|(x^+(A)-x^+(I))(x^-(A)-x^-(I))\right|^2 \,.
\end{equation}
The expressions above depend on the coordinate system, 
which should be chosen so that $t_\pm=0$. 
Here, the anchor curve is located far outside the black hole. 
The position of the island is determined so that the generalized entropy is extremized, 
and then, the island is placed inside the event horizon as we will see below. 

\begin{figure}[H]
\centering
\includegraphics[scale=0.5]{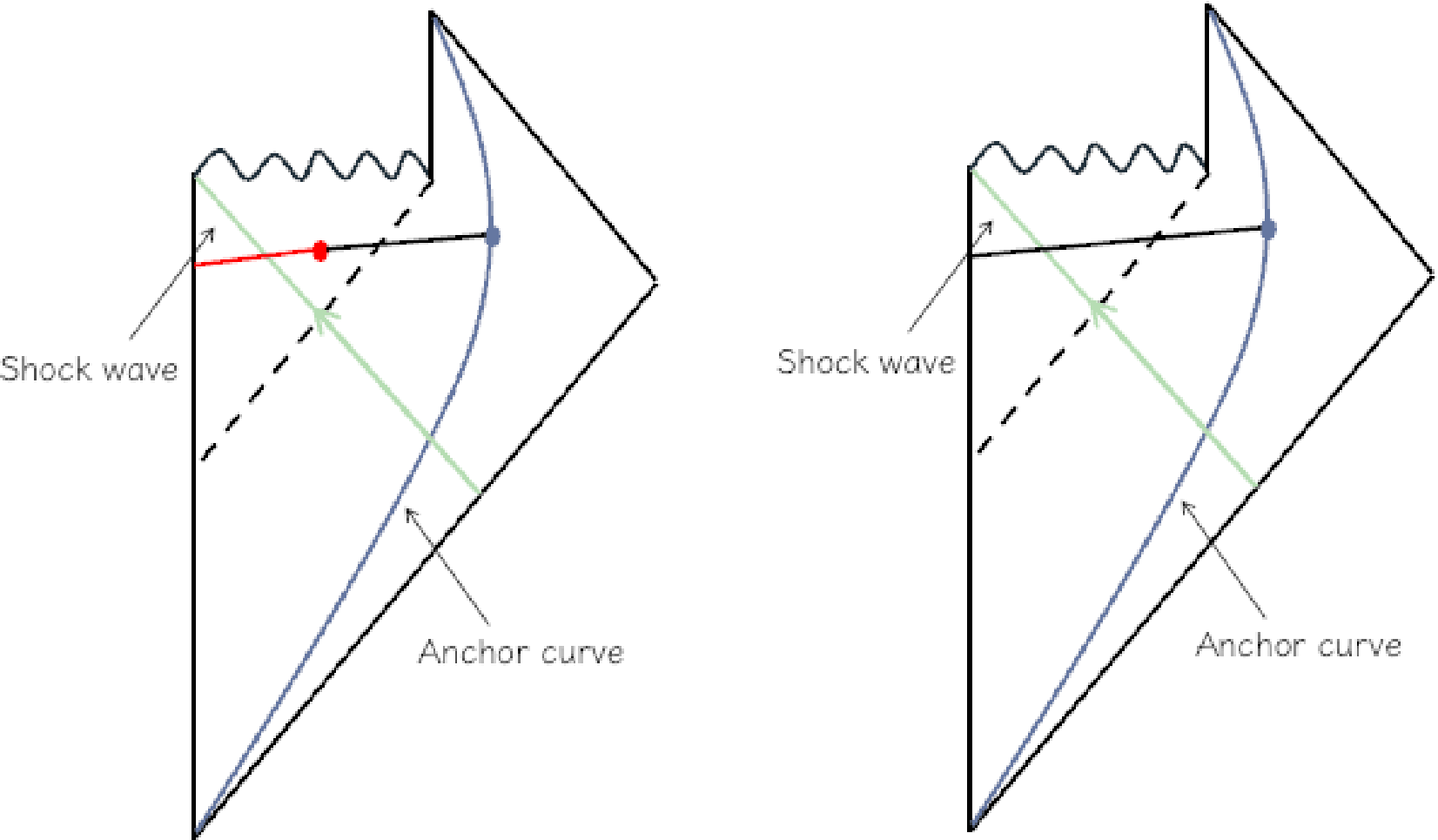}
\caption{This is the Penrose diagram of the RST black hole. The red line represents the island and the blue line the anchor curve.}
\label{fig-QFC-RST}
\end{figure}

In the absence of islands, the generalized entropy is given by
\begin{align}
S_{\text{gen}} = 
\frac{\text{Area(A)}}{4G_{N}} + S_{\text{bulk}}[\mathcal{S}_{\text{A0}}] \,. 
  \end{align}
The surface $\mathcal{S}_{A0}$ extends 
from the anchor curve to the inner boundary of 
the spacetime defined by $\Omega=\Omega_\text{crit}$. 
The situation we are considering is illustrated by Figure~\ref{fig-QFC-RST}.

\subsubsection{Island}
The generalized entropy in the presence of an island is expressed as
\begin{align}
S^{\text{island}}_{\text{gen}} = 
&2M \left\{ 1-v_{\text{I}}(1+u_{\text{I}}) - \epsilon\log(-Mv_{\text{I}}u_{\text{I}} )  \right\} 
\nn\\
&+\frac{c}{12}\log \left[ 
                                 \Big(\log\frac{v_{\text{A}}}{v_{\text{I}}}\log\frac{u_{\text{A}}}{u_{\text{I}}}\Big)^2 \frac{v_{\text{A}}u_{\text{A}}}{1-v_{\text{A}}(1+u_{\text{A}})}
\frac{v_{\text{I}}u_{\text{I}}}{1-v_{\text{I}}(1+u_{\text{I}})} \right] \,. 
\label{island gen}
\end{align}
To find the position of the island $(v_{I},u_{I})$, we extremize \eqref{island gen} over $(v_{I},u_{I})$ and obtain the following two conditions,
\begin{align}
0 &= -2M(1+u_{I}) + \dfrac{c}{12 v_{I}} \frac{1}{1-v_{I}(1+u_{I})} - \dfrac{c}{24v_{I}} 
       - \dfrac{c}{6v_{I}{\log \left( \dfrac{v_{A}}{v_{I}} \right) }} \,, 
\label{vi derivative}
\\
0&= -2Mv_{I} + \dfrac{c}{12v_{I}} \dfrac{1-v_{I}}{1-v_{I}(1+u_{I})} - \dfrac{c}{24u_{I}} - \dfrac{c}{6v_{I} \log \left(\frac{u_{A}}{u_{I}} \right) } \,.
\label{ui derivative}
\end{align}
Here, we can drop the last term on the right side of \eqref{vi derivative}, when $\log ({v_{A}}/{v_{I}})\gg 1$. Let us see that this is indeed the case. Using \eqref{vi derivative} and \eqref{ui derivative}, we obtain  the following equations:
\begin{align}
(v_{I}(1+u_{I}))^2 - (1-\epsilon) v_{I}(1+u_{I}) + \epsilon = 0 \,, \quad  
\log\frac{u_{A}}{u_{I}} = 4(1+u_{I}) \,. 
\label{equation1}
\end{align}
Solving \eqref{equation1} for $v_{I}(1+u_{I})$, we obtain 
\begin{align}
u_{I} = -1 + \frac{\epsilon}{v_{I}} + O(\epsilon^2).
\label{solve1}
\end{align}
The other solution is unphysical near the singularity. Substituting the island solution \eqref{solve1} into \eqref{equation1}, we have
\begin{align}
 v_{I}= -\frac{3\epsilon}{(u_{A}+1)} 
+ O(\epsilon^2) \,.
\label{solve2}
\end{align}
Next, introducing
\begin{align}
u_{A}= -1 - e^{\tilde{\sigma}_{A}-t_{A}} \,, \qquad v_{A} = e^{\tilde{\sigma}_{A}+t_{A}} \,,
\label{asymptotic coordinate}
\end{align}
and using \eqref{solve2} and \eqref{asymptotic coordinate}, we obtain
\begin{align}
\log \frac{v_{A}}{v_{I}} \sim 2\tilde{\sigma}_{A} + \log\frac{1}{\epsilon} \gg 1\,. 
\end{align}
Thus, $\log ({v_{A}}/{v_{I}}) \gg 1$ holds and the last term of \eqref{vi derivative} can be dropped. Finally, substituting \eqref{solve1} into \eqref{island gen}, we obtain
\begin{align}
S^{\text{island}}_{\text{gen}} = 2M -\frac{c}{24}(t_{A}-\tilde{\sigma}_{A})+ \frac{c}{6} \tilde{\sigma}_{A}+ \cdots \,. 
\end{align}

\subsubsection{No-island}
We consider the generalized entropy $S^{\text{no-island}}_{\text{gen}}$ in the absence of an island. 
Let the surface $\mathcal S_{AI}$ extend to a fixed reference point $(v_{0}, u_{0})$ 
on the boundary $\Omega=\Omega_\text{crit}$ 
in the region $v<1$: i.e., $v_{0} < 1$ and $u_{0} = -\epsilon/v_{0}$. 
Then, we can express the generalized entropy $S^{\text{no-island}}_{\text{gen}}$ as 
\begin{align}
S^{\text{no-island}}_{\text{gen}} &= 
\frac{c}{12}\log \left[ \left(\log \dfrac{v_{\text{A}}}{v_{\text{0}}}\log \dfrac{u_{\text{A}}}{u_{\text{0}}} \right)^2 \dfrac{v_{\text{A}}u_{\text{A}}}{1-v_{\text{A}}(1+u_{\text{A}})} \right]
\nn\\
&=\frac{c}{12}(t_{A}-\tilde{\sigma}_{A}) + \frac{c}{6}\tilde{\sigma}_{A} + \cdots \,.
\end{align}

\section{The quantum focusing conjecture in RST model}\label{qfc}
In this section, we show that the QFC holds for two-dimensional evaporating black holes in the RST model.

As expected from \eqref{def:QFC} \eqref{def:S_gen} \eqref{area}, 
let us consider the {\em quantum expansion in two-dimensions} \footnote{
More precisely we choose the correspondence as ${\cal A}/(4G_N) = 2\Omega$.
}: 
\begin{align}
\label{def:2DQexp}
 \Theta := \frac{1}{2 \Omega } \frac{dS_{\text{gen}}}{d\lambda} \,. 
\end{align}
Note that one can easily check that if $\Theta$ is defined without $\Omega$ in the denominator, the QFC for such defined $\Theta$ fails to hold, except for some circumstances [see, e.g., Case (ii) below]. For more details on a subtle issue in the definition of the quantum expansion, see Appendix A.  

As the generalized entropy $S_{\text{gen}}$ in the above definition \eqref{def:2DQexp}, we consider two expressions: one  $S^{\text{island}}_{\text{gen}}$ for island configuration and the other $S^{\text{no-island}}_{\text{gen}}$ for no island configuration, which we reviewed in the previous section. 
Restoring the area term at the anchor point, 
they are expressed as follows:
\begin{align}
S^{\text{island}}_{\text{gen}} = 
&2M \left\{1-v_{\text{I}}(1+u_{\text{I}}) - \epsilon\log(-Mv_{\text{I}}u_{\text{I}})\right\}
\nn\\
&+2M \left\{1-v_{\text{A}}(1+u_{\text{A}}) - \epsilon\log(-Mv_{\text{A}}u_{\text{A}})\right\}
\nn\\
&+\frac{c}{12}\log\left[ \left(\log\frac{v_{\text{A}}}{v_{\text{I}}}\log\frac{u_{\text{A}}}{u_{\text{I}}} \right)^2 \frac{v_{\text{A}}u_{\text{A}}}{1-v_{\text{A}}(1+u_{\text{A}})}
\frac{v_{\text{I}}u_{\text{I}}}{1-v_{\text{I}}(1+u_{\text{I}})}
\right] \,, 
\label{island gen study}\\
S^{\text{no-island}}_{\text{gen}} = 
&2M \left\{1-v_{\text{A}}(1+u_{\text{A}}) - \epsilon\log(-Mv_{\text{A}}u_{\text{A}})\right\}
\nn\\
&+\frac{c}{12}\log\left[\left(\log\frac{v_{\text{A}}}{v_{\text{0}}}\log\frac{u_{\text{A}}}{u_{\text{0}}}\right)^2 \frac{v_{\text{A}}u_{\text{A}}}{1-v_{\text{A}}(1+u_{\text{A}})} \right] \,,
\label{no-island gen study}
\end{align}
where the position of (the boundary of) the island $(u_I,v_I)$ are given by \eqref{solve1}--\eqref{solve2}, 
as we have seen in the previous section. 
Note that the area of the anchor curve is ignored in \cite{Gautason:2020tmk} 
as it is an irrelevant constant for the Page curve, 
while it plays an important role in the quantum focusing conjecture.

We examine the behavior of these two entropies, $S^{\text{island}}_{\text{gen}}$ and $S^{\text{no-island}}_{\text{gen}}$, in the following three cases (i)~near the horizon $\epsilon \ll -v_{\text{A}}(1+u_{\text{A}}) \ll 1$ ,
(ii)~away from the horizon $-v_{A}(1+u_{A})=O(1)$  and (iii)~very near the horizon $-v_{A}(1+u_{A})\sim \epsilon$ , separately. Our goal is to show, for each case of (i)--(iii), 
\begin{align}
 \dfrac{\partial}{\partial v} \left[ \left( \dfrac{e^{-2\rho}}{\Omega} \dfrac{\partial }{\partial v} S_{\rm gen}^{\text{island/no-island}}\right) \right] \leq 0 \,. 
\end{align}

\subsection{Case (i): $\epsilon \ll -v_{\text{A}}(1+u_{\text{A}}) \ll 1$ (near the horizon)}

In this case, 
$\Omega$ can be approximated as 
\begin{align}
 \Omega(A) 
 &\simeq 
 M - \frac{c}{48}\log v_A \,,
 \label{OmegaA1}
 \\
 \Omega(I) 
 &\simeq 
 M + \frac{c}{48}\log(-1-u_A) \,, 
 \label{OmegaI1}
\end{align}
and the bulk entropy $S_\text{bulk}$ is 
\begin{equation}
 S_\text{bulk}^\text{island}  
 \simeq 
 \frac{c}{12}\log v_A + \frac{c}{12}\log(-1-u_A) \,,
\end{equation}
for the island configuration and 
\begin{equation}
 S_\text{bulk}^\text{no-island} 
 \simeq 
 \frac{c}{12}\log v_A \,,
\end{equation}
for the no-island configuration. 
Using these relations, 
we can express $S^{\text{island}}_{\text{gen}}, \,S^{\text{no-island}}_{\text{gen}}$ approximately as
\begin{align}
S^{\text{island}}_{\text{gen}} &\simeq 4M + \frac{c}{8}\log(-u_{A}-1)+\frac{c}{24}\log v_{A} \,, 
\label{sim island gen1}
\\
S^{\text{no-island}}_{\text{gen}} & \simeq 2M + \frac{c}{24}\log v_{A} \,. 
\label{sim no-island gen1}
\end{align} 

\subsubsection{GSL} 
Taking the derivative with respect to $v_{A}$, we find
\begin{align}
\del_{v_{\text{A}}}S^{\text{island}}_{\text{gen}} = \frac{c}{24} \frac{1}{v_{A}} \,, 
\\
\del_{v_{\text{A}}}S^{\text{no-island}}_{\text{gen}} = \frac{c}{24}\frac{1}{v_{A}} \,, 
\end{align}
both of which are non-negative. Therefore both $S_{\text{gen}}^{\text{island}}$ and $S_{\text{gen}}^{\text{no-island}}$ are non-decreasing. 

\subsubsection{QFC} 
Next, to see if the QFC holds, we check $\del^2_{\lambda}S_{\text{gen}}\leq 0$. 
We have 
\begin{align}
\del^2_{\lambda} S_{\text{gen}} = e^{-4\rho} (\del^2_{+} S - 2\del_{+}\rho \del_{+}S) \,.
\end{align}
Since 
$\Omega = e^{-2\rho} + {c} \rho/{24} = M + O(\epsilon)$ as we have seen in \eqref{OmegaA1}, 
it follows that 
$\partial_+\rho = O(\epsilon)$. 
As we have also found in \eqref{sim island gen1}--\eqref{sim no-island gen1} 
that $\partial_+ S_\text{gen} = O(\epsilon)$, 
the second term is zero 
up to $O(\epsilon^2)$ terms. 
The first term is calculated as  
\begin{align}
\del^2_{+} S^{\text{island}}_{\text{gen}} = -\Big(\frac{\del v_{\text{A}}}{\del x^{+}} \Big)^2\frac{c}{24} \frac{1}{v^2_{\text{A}}} <0 \,,
\\
\del^2_{+} S^{\text{no-island}}_{\text{gen}} = -\Big(\frac{\del v_{\text{A}}}{\del x^{+}}\Big)^2\frac{c}{24} \frac{1}{v^2_{\text{A}}}<0 \,.
\end{align}
Thus, the QFC is satisfied,
\begin{align}
\frac{d\Theta}{d\lambda} <0 \,.
\end{align}

\subsection{Case (ii):$-v_{A}(1+u_{A})=O(1)$(away from the horizon)}

This case implies that $v_{\text{A}}$ and $v_{\text{I}}$ are far away from each other. 
By using 
\eqref{solve1}--\eqref{solve2}, 
the generalized entropy can be approximately expressed as
\begin{align}
S^{\text{island}}_{\text{gen}} \simeq \,\,&4M 
+ \frac{c}{24} 
+ \frac{c}{24}\log (u_{\text{A} }+ 1)
-\frac{c}{24} \log \left[ -1-\dfrac{1}{3}(u_{\text{A} }+ 1) \right] 
\nn \\
&+ \frac{c}{6} \log \left[ \log \frac{-v_{\text{A}}(u_{\text{A}}+1)}{3\epsilon} \right] 
+\frac{c}{6}\log \left( \log \frac{-3u_{\text{A}}}{4+u_{\text{A}}} \right)  
+\frac{c}{24}\log u_{\text{A}} 
\nn \\
&- \frac{c}{12} \log\left[1-v_{\text{A}}(1+u_{\text{A}}) \right]
-\frac{c}{12} \log(-u_{\text{A}}-1)
+\frac{c}{12}\log (-4-u_{\text{A}})
\nn\\
&-2M v_{\text{A}}(1+u_{\text{A}})
+\frac{c}{24} \log v_{\text{A}} \,, 
\label{sim island gen2}
\\
S^{\text{no-island}}_{\text{gen}} \simeq
&\,\,2M
-2M v_{\text{A}}(1+u_{\text{A}})
+\frac{c}{24}\log v_{\text{A}}
+\frac{c}{24}\log u_{\text{A}}\nn\\
&+\frac{c}{6}\log \left( \log \frac{v_{\text{A}}}{v_{\text{0}}} \right)
+\frac{c}{6}\log \left( \log \frac{u_{\text{A}}}{u_{\text{0}}} \right)
- \frac{c}{12}\log \left[ 1-v_{\text{A}}(1+u_{\text{A}}) \right] \,.
\label{sim no-island gen2}
\end{align}

\subsubsection{GSL} 
Taking the derivative of \eqref{sim island gen2} and \eqref{sim no-island gen2} with respect to $v_{A}$, we find,
\begin{align}
&\del_{v_{\text{A}}}S^{\text{island}}_{\text{gen}} = 
 \dfrac{c}{6} \dfrac{1}{\log \left[\dfrac{-v_{\text{A}}(u_{\text{A}}+1)}{3\epsilon} \right]} \dfrac{1}{v_{\text{A}}}
+\dfrac{c}{12} \dfrac{1+u_{\text{A}}}{1-v_{\text{A}}(1+u_{\text{A}})} 
-2M(1+u_{\text{A}})
+\dfrac{c}{24} \dfrac{1}{v_{\text{A}}} \,,
\label{island positive}
\\
&\del_{v_{\text{A}}}S^{\text{no-island}}_{\text{gen}} = 
 \dfrac{c}{6} \dfrac{1}{ \log \left(\dfrac{v_{\text{A}}}{v_{\text{0}}} \right)} \dfrac{1}{v_{\text{A}}}
+\dfrac{c}{12} \dfrac{1+u_{\text{A}}}{1-v_{\text{A}}(1+u_{\text{A}})}
-2M(1+u_{\text{A}}) 
+ \dfrac{c}{24} \dfrac{1}{v_{\text{A}}} \,. 
\label{no-island positive}
\end{align}
Let us consider \eqref{island positive}, first. 
The first term in the right-hand side is positive since $-v_{A}(1+u_{A})$ is assumed to be $O(1)$. 
The fourth term is obviously positive as $v_A>0$. 
Now, we 
compare the second and third terms. Because  $1-v_{A}(1+u_{A})$ is greater than 1, we immediately find
\begin{align}
-(1+u_{\text{A}}) >-\frac{1+u_{\text{A}}}{1-v_{\text{A}}(1+u_{\text{A}})} \,.
\label{2.2.1 compare}
\end{align} 
As we have assumed $\epsilon = {c}/({48M}) \ll 1$, we obtain the following relations: 
\begin{align}
-2M(1+u_{\text{A}}) >-\frac{c}{12}(1+u_{\text{A}}) >-\frac{c}{12} \frac{1+u_{\text{A}}}{1-v_{\text{A}}(1+u_{\text{A}})}\,.
\end{align}
This implies that the right-hand side of \eqref{island positive} is positive definite. 
For eq.~\eqref{no-island positive}, the first term is positive because $v_A>v_0$. Then, the same argument to \eqref{island positive} can be applied to the other terms in eq.~\eqref{no-island positive}. Therefore both $S^{\text{island}}_{\text{gen}}$ and $S^{\text{no-island}}_{\text{gen}}$ are non-decreasing.

\subsubsection{QFC}
\begin{figure}[H]
\begin{center}
\includegraphics[scale=0.6]{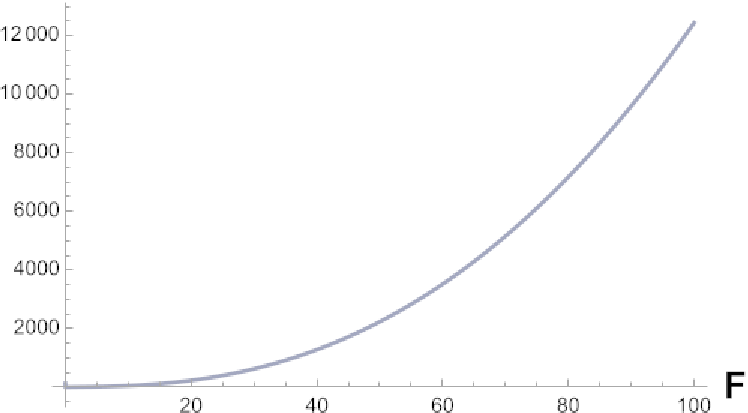}
\caption{The diagram depicts the numerator (inside the brackets) of \eqref{fig-QFC numerator}. It is positive, and hence the right-hand side of \eqref{fig-QFC numerator} as a whole is negative. Here, the horizontal axis is F$=-v(u+1)/3\epsilon$.}
\label{fig-QFC}
\end{center}
\end{figure}

Let us check the QFC:
\begin{align}
\del_{v}\Big(\frac{e^{-2\rho}}{\Omega}\del_{v}S \Big) \leq 0 \,. 
\label{QFC2}
\end{align}
For simplicity, in the following we consider in the first order of $\epsilon$ and ignore the higher order terms. First, we note that the factor inside the parentheses of \eqref{QFC2} can be estimated as follows:  
\begin{align}
\frac{e^{-2\rho}}{\Omega} 
&\simeq 
1+ \frac{c}{48\Omega}\log \Omega
\nn\\
&= 1+ \epsilon \frac{\log M(1-v(u+1)-\epsilon \log(-Muv))}{1-v(u+1)-\epsilon \log(-Muv)}
\nn\\
&\simeq 1+  \frac{\epsilon \log M \left\{1-v(u+1) \right\}}{1-v(u+1)} \,.
\label{QFC1}
\end{align}
Then, for the generalized entropy for island, we find
\begin{align}
\left(\del_{v} \dfrac{e^{-2\rho}}{\Omega} \right) 
\del_{v}S^{\text{island}}_{\text{gen}}
= -\dfrac{c}{24} \dfrac{(u+1)^2}{\left[1-v(u+1)\right]^2} \log \left\{ M(1-v(u+1))-1 \right\} \,,  
\end{align}
and also
\begin{align}
\del_{v}^2 S^{\text{island}}_{\text{gen}}
= &- \dfrac{c}{6} \dfrac{1}{v^2} \dfrac{1}{\log \left[-\dfrac{v(u+1)}{3\epsilon} \right]}
-\dfrac{c}{6} \dfrac{1}{v^2} \dfrac{1}{\left\{\log \left[-\dfrac{v(u+1)}{3\epsilon} \right] \right\}^2}
\nn\\
&+ \dfrac{c}{12}\left[\dfrac{u+1}{(1-v(u+1))} \right]^2
-\dfrac{c}{24} \dfrac{1}{v^2} \,. 
\end{align}
Therefore we have,
\begin{align}
&\del_{v}\left( \dfrac{e^{-2\rho}}{\Omega}\del_{v}S \right)=
\left(\del_{v} \dfrac{e^{-2\rho}}{\Omega} \right) 
\del_{v}S^{\text{island}}_{\text{gen}}
+ \del_{v}^2 S^{\text{island}}_{\text{gen}}
\nn\\
&=-\dfrac{c}{24}
\dfrac{1}{v^2[1-v(u+1)]^2 \left\{ \log \left[-\dfrac{v(u+1)}{3\epsilon} \right] \right\}^2 }
 \cdot 
\nn \\
& \quad \times 
\Bigg[ 
  \left\{1-v(u+1) \right\}^2 \left\{ 2+ \log \left( -\dfrac{v(u+1)}{3\epsilon} \right) \right\}^2 
\nn \\
& \qquad \quad 
 + \left\{ \log \left[-\dfrac{v(u+1)}{3\epsilon} \right] \right\}^2 (1+u)^2v^2 \left\{\log M \left[1-v(u+1) \right]-3 \right\}
\Bigg] \,. 
\label{fig-QFC numerator}
\end{align}
If the above equation is negative, then the QFC is satisfied. This is valid if $M$ is sufficiently large. We can immediately find that the right-hand side of \eqref{fig-QFC numerator} is non-negative, as can be seen in Figure\ref{fig-QFC}, which shows the positivity of the numerator of the right-hand side of \eqref{fig-QFC numerator}. 
Thus, the QFC holds. 

Similarly, we check that the QFC is satisfied in the case of no-island. We find
\begin{align}
\del_{v}\left( \dfrac{e^{-2\rho}}{\Omega} \right) \del_{v}S^{\text{no-island}}_{\text{gen}}
= -\dfrac{c}{24}\frac{(u+1)^2}{[1-v(u+1)]^2} \log\{M[1-v(u+1)] -1 \} \,. 
\end{align}
We can also calculate $\del_{v}^2 S^{\text{no-island}}_{\text{gen}}$ as 
\begin{align}
\del_{v}^2 S^{\text{no-island}}_{\text{gen}}
= &-\dfrac{c}{6}\dfrac{1}{v^2} \dfrac{1}{\log \left(\dfrac{v}{v_{0}} \right) }
  -\dfrac{c}{6 }\dfrac{1}{v^2} \dfrac{1}{\left\{ \log \left( \dfrac{v}{v_{0}} \right) \right\}^2}
\nn\\
&+ \dfrac{c}{12} \left\{ \dfrac{u+1}{[ 1-v(u+1) ]} \right\}^2
-\dfrac{c}{24}\dfrac{1}{v^2} \,. 
\end{align}
Therefore,
\begin{align}
&\del_{v}\left( \dfrac{e^{-2\rho}}{\Omega}\right) \del_{v}S^{\text{no-island}}_{\text{gen}}
+ \del_{v}^2 S^{\text{no-island}}_{\text{gen}}
\nn\\
&=-\frac{c}{24} \dfrac{1}{v^2[1-v(u+1)]^2[\log(-\dfrac{v}{v_{0}})]^2 } \cdot 
\nn \\
& \quad 
  \times \Bigg[
[1-v(u+1)]^2 \left\{ 2 + \log \left( \dfrac{v}{v_{0}} \right) \right\}^2 
\nn \\
& \qquad \quad 
  + \left[ \log \left(\dfrac{v}{v_{0}} \right) \right]^2 (1+u)^2 v^2 \left\{ \log [M(1-v(u+1)]-3\right\} 
            \Bigg] \,. 
\label{no-island QFC2}
\end{align}
This has much the same form as in the case of island. By the same argument before, we find that the right-hand side of \eqref{no-island QFC2} is negative.
Thus, the QFC is satisfied.

\subsection{Case (iii):$-v_{A}(1+u_{A})\sim \epsilon$ (very near the horizon) }

Finally, we discuss the case $-v_{A}(1+u_{A})\sim \epsilon $, in which $v_{A}$ and $v_{I}$ are very close. In this case, the assumption $\log (v_{A}/{v_{I}})\gg 1$ no longer holds, and the approximation used in \eqref{solve1} cannot be applied. 
Although the position of (the boundary of) the island is determined 
so that the generalized entropy is extremized, 
we do not substitute the position explicitly 
but treat $u_I$ and $v_I$ as functions of $u_A$ and $v_A$. 

\subsubsection{GSL}
We calculate $\del_{{v}_{A}}S_{\text{gen}}^{\text{island}}$ using generalized entropy \eqref{island gen study}.
\begin{align}
\del_{{v}_{A}}S_{\text{gen}}^{\text{island}}
= \dfrac{c}{6v_{A}}\dfrac{1}{\log \left(\dfrac{v_{A}}{v_{I}}\right)}
+\dfrac{c}{12}\dfrac{1+u_{A}}{ \left[ 1-v_{A}(1+u_{A}) \right]} 
-2M(1+u_{A}) +\dfrac{c}{24} \dfrac{1}{v_{A}} \,. 
\label{va derivative3}
\end{align}
By the same argument before, we find that the right-hand side of \eqref{va derivative3} is positive-definite, and therefore that $S_{\text{gen}}^{\text{island}}$ is non-decreasing.

\subsubsection{QFC}
To see whether QFC holds, we consider the following second-order derivative,
\begin{align}
&\del_{v_{A}} \left( \dfrac{e^{-2\rho}}{\Omega} \del_{v_{A} }S^{\text{island}}_{\text{gen}}
                \right)
\nn\\
&=\dfrac{\del}{\del v_{A}} \left( \dfrac{e^{-2\rho}}{\Omega} \del_{v_{A} }S^{\text{island}}_{\text{gen}}\right)
+ \dfrac{\del v_{I}}{\del v_{A}}\dfrac{\del}{\del v_{I}} \left(\dfrac{e^{-2\rho}}{\Omega} \del_{v_{A} }S^{\text{island}}_{\text{gen}}\right)
+ \dfrac{\del u_{I}}{\del v_{A}}\dfrac{\del}{\del u_{I}} \left( \dfrac{e^{-2\rho}}{\Omega} \del_{v_{A} }S^{\text{island}}_{\text{gen}}\right),
\nn\\
&=\dfrac{\del}{\del v_{A}} \left(\dfrac{e^{-2\rho}}{\Omega} \del_{v_{A} }S^{\text{island}}_{\text{gen}}\right) 
+ \dfrac{\del v_{I}}{\del v_{A}} \dfrac{c}{6v_{A}v_{I}\left(\log \dfrac{v_{A}}{v_{I}}\right)^2} \,, 
\label{QFC3}
\end{align}
where the position of (the boundary of) the island $(u_I,v_I)$ 
depends on the position of the anchor point $(u_A,v_A)$. 
The partial derivative $\partial_{v_A}$ in the first line includes the variation of $v_A$ 
in $u_I(u_A,v_A)$ and $v_I(u_A,v_A)$, 
while those in the second and third line $\partial/\partial v_A$ is for fixed $(u_I,v_I)$. 
The third term in the second line is calculated to be zero. The first term in the third line is found to be negative by the same calculation as before. Therefore, in order for QFC to hold, $\del v_{I}/\del v_{A}$ must be negative.

Let us check whether $\del v_{I}/\del v_{A}$ is negative. First we note that eqs.~\eqref{vi derivative} and \eqref{ui derivative} are rewritten as,
\begin{align}
0 &= -2M(1+u_{I}) + \dfrac{c}{12} \frac{1+u_{I}}{1-v_{I}(1+u_{I})} + \dfrac{c}{24v_{I}} 
- \dfrac{c}{6v_{I}{\log \left(\dfrac{v_{A}}{v_{I}} \right)}} \,, 
\label{vi derivative2}
\\
0&= -2Mv_{I} + \dfrac{c}{12} \dfrac{v_{I}}{1-v_{I}(1+u_{I})} +\dfrac{c}{24u_{I}} - \dfrac{c}{6u_{I}\log\left(\dfrac{u_{A}}{u_{I}}\right)} \,.
\label{ui derivative2}
\end{align}
Noting that $v_{I} \gg 1,u_{I}\sim -1$, from eqs. \eqref{vi derivative2} and \eqref{ui derivative2}, we obtain
\begin{align}
0  &\simeq \dfrac{c}{-12M v_{I}(1+u_{I})+{c}/{4}} + \log \dfrac{v_{A}}{v_{I}} \,, 
\label{vi derivative sim2}
\\
0 &\simeq -\dfrac{c}{12M v_{I}} + (u_{I}-v_{A}) \,. 
\label{ui derivative sim2}
\end{align}
Taking 
${v_{A}}$-derivatives of 
\eqref{vi derivative sim2} and \eqref{ui derivative sim2}, we have respectively
\begin{align}
0 &= \frac{d}{dv_{A}}\frac{\del S}{\del v_{I}} 
\nn \\ 
 &= \frac{1}{v_{A}} 
- \frac{\del v_{I}}{\del v_{A}} \left[ 
                                              \dfrac{12Mc(1+u_{I}) }{ \left\{-12Mv_{I}(1+u_{I})+ {c}/{4} \right\}^2}+ \dfrac{1}{v_{I}}
                                       \right] 
- \dfrac{\del u_{I}}{\del v_{A}} \dfrac{12Mcv_{I}}{\left\{-12Mv_{I}(1+u_{I})+{c}/{4} \right\}^2} \,, 
\label{vavi derivative1}
\\
0&= \dfrac{d}{dv_{A}}\dfrac{\del S}{\del u_{I}}=
\dfrac{\del v_{I}}{\del v_{A}} \dfrac{c}{12Mv_{I}^2} 
+ \dfrac{\del u_{I}}{\del v_{A}} \,. 
\label{vaui derivative1}
\end{align}
From \eqref{vavi derivative1} and \eqref{vaui derivative1}, we obtain 
\begin{align}
0&=  \dfrac{\left\{-12Mv_{I}(1+u_{I})+{c}/{4} \right\}^2}{12Mcv_{I}v_{A}} 
- \dfrac{\del v_{I}}{\del v_{A}} \left[\frac{1+u_{I}}{v_{I}}
+ \dfrac{\left\{-12Mv_{I}(1+u_{I})+{c}/{4} \right\}^2}{12Mcv_{I}^2} -  \dfrac{c}{12Mv_{I}^2} \right]
\nn\\
&\equiv A -\frac{\del v_{I}}{\del v_{A}}B \,. 
\end{align}
As mentioned below eq.~\eqref{QFC3}, in order for QFC to hold, $\del v_{I}/\del v_{A}$ needs to be negative. Now we will show that $B$ is negative, as $A$ is obviously positive. $B$ is written as, 
\begin{align}
B=
\frac{\left\{12Mv_{I}(1+u_{I}) \right\}^2 + 6Mcv_{I}(1+u_{I}) - {15c^2}/{16}}{12Mcv_{I}^2} \,. 
\label{define b}
\end{align}
Defining $X= {12M}v_{I}(1+u_{I})/c$, we can express the numerator of \eqref{define b} as $X^2 + X/2 -{15}/{16} $. 
Thus, the numerator of \eqref{define b} vanishes when $X=-5/4, 3/4$, 
and $B$ is negative for $-5/4 < X < 3/4$. 

\begin{figure}[H]
\begin{center}
\includegraphics[scale=0.6]{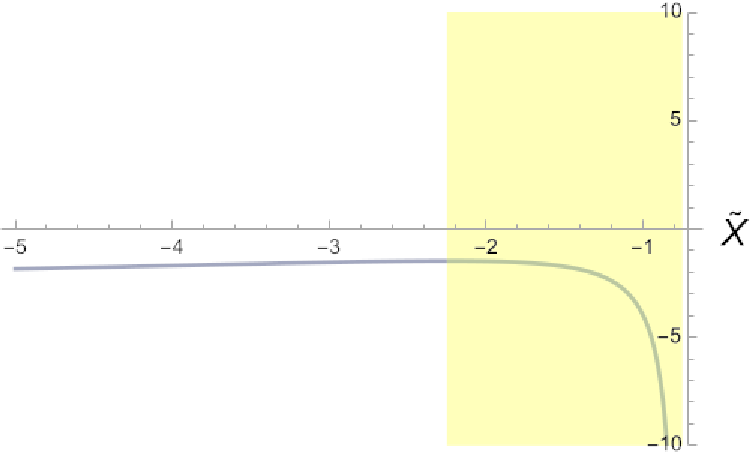}
\caption{This figure depicts the solution of eq.~\eqref{negative true}. The vertical axis is $-\log v_A$ and the horizontal axis ${\tilde X}=12Mv_I(1+u_A)/c$. From eq.~\eqref{negative true}, one finds that an island solution exists in the yellow region, where $-9/4< {\tilde X}< -3/4$.} \label{negative QFC}
\end{center}
\end{figure}

Now, we will show that an island exists only for $-5/4 < X < 3/4$. 
The position of (the boundary of) the island is 
determined by \eqref{vi derivative sim2}--\eqref{ui derivative sim2}. 
Note that \eqref{ui derivative sim2} can be transformed as follows:
\begin{align}
u_{I} = \frac{c}{12Mv_{I}} + u_{A} \,. 
\end{align}
We substitute this in \eqref{vi derivative sim2} and obtain 
\begin{align}
\dfrac{c}{{12Mv_{I}} (1+u_{A})+ {3c}/{4}} -\log v_{I} = - \log v_{A} \,.
\label{negative true}
\end{align}
Since $v_A\gg 1$, the right-hand side is negative. 
Eq.~\eqref{negative true} has two solutions for sufficiently large $v_A$, 
but has no solution if $v_A$ is too small (see Figure~\ref{negative QFC}). 
In $v_A\to\infty$, two solutions are $v_I \simeq - {3\epsilon}/({u_A+1})$ and $v_I\simeq v_A$. 
The former is the position of the island, which we have seen in the previous section, \eqref{solve2}, 
and the latter gives a larger value of the generalized entropy, which is a false saddle point. 
Two solutions approach each other as $v_A$ is lowered, and eventually merges at some critical point, 
where the derivative of \eqref{negative true} vanishes.

Taking the derivative of \eqref{negative true} with respect to $v_{I}$ and then multiplying by $v_{I}$ we obtain
\begin{align}
\left\{ \frac{12Mv_{I}}{c} (1+u_{A})+\frac{3}{4} \right\}^2+\frac{12M}{c}v_{I} (1+u_{A})=0.
\label{equationdifferent}
\end{align}
Defining $\tilde{X} = {12M}v_{I} (1+u_{A})/c$, we can express the above equation \eqref{equationdifferent} as $\tilde{X} + (\tilde{X}+{3}/{4})^2 = 0 $. Thus, the solutions to \eqref{equationdifferent} are $\tilde{X}=-1/4, -9/4$. 
This implies that two solutions of \eqref{negative true} are located in 
$\tilde X < -9/4$ and $-9/4 < \tilde X < -3/4$, respectively. 
The solution in $-9/4 < \tilde X < -3/4$ gives the position of the island and that in $\tilde X < -9/4$ is the false saddle. 
Note that $\tilde{X}=X+1$ and therefore an island exists when 
$-9/4 < \tilde{X} < -3/4$, that is when $-5/4 < X < 1/4$. 
As seen above, when $-5/4 < X < 3/4$, $\partial v_I/ \partial v_A < 0$, 
and therefore 
\begin{align}
\frac{d\Theta}{d\lambda} <0.
\end{align}
Thus, the QFC is satisfied.

\section{Conclusion}\label{conclusion}
In this paper, we have shown that the QFC holds for a two-dimensional evaporating black hole in RST model, with quantum backreaction on the geometry as well as the island formation taken into account. 
We have given a suitable definition of the quantum expansion in two-dimensions. 
The generalized entropy $S_\text{gen}$ is defined by the sum of the area entropy and the entropy of matter fields. 
In two dimensions, a Cauchy surface is separated by a point, and hence there is no notion of the area of that point. 
We treated $\Omega$ as the area entropy as in the previous studies on the RST model. 
In this paper, we showed that the generalized entropy is non-decreasing along outgoing null lines, 
even if the island affects the entropy of matter fields. 
We defined the quantum expansion \eqref{def:2DQexp}, 
$\Theta = \dfrac{1}{2\Omega} \dfrac{dS_\text{gen}}{d \lambda}$, 
as the quantum expansion is defined as a variation of the generalized entropy per unit volume. 
Then, the QFC with our definition of the quantum expansion is satisfied for evaporating black holes in the RST model, 
in both cases of the island configuration and no-island configuration. 

A most important advantage of the RST model is that 
solutions including quantum effects can be obtained exactly. 
In the previous work \cite{Matsuo:2023cmb}, one of the authors 
studied the four-dimensional Schwarzschild black holes, for which  
the solution cannot be obtained exactly. The solution was approximated by 
the Vaidya metric, and only most important part 
of quantum effects was taken into calculation as an approximation.
In the RST model, quantum effects of matter are fully taken into account, 
and hence, our result implies that the QFC cannot be violated by ignored quantum effects. 
Moreover, the RST model agrees with the Einstein gravity 
in large numbers of dimensions \cite{Emparan:2013xia,Soda:1993xc}. 
By focusing only on the spherically symmetric configurations,
$D$-dimesnional Einstein gravity in the large-$D$ limit reduces to the two-dimensional dilaton gravity. 
The Schwarzschild black hole in the large-$D$ limit is equivalent to 
the eternal black hole in the CGHS model, or equivalently in the classical limit of the RST model.  
Thus, the result of our paper is expected to be true also for the Schwarzschild black holes. 
Although the QFC in general black holes is left to be checked, 
our result is not only for the RST model itself, 
but also related to a more realistic model of the Einstein gravity.

In higher-dimensions, the QFC was used to derive the QNEC ~\cite{Bousso:2015wca,Koeller:2015qmn,Fu:2017evt}. 
In the two dimensional case, since the generalized entropy is given in terms of $\Omega$ as: 
\begin{align}
S_{\text{gen}} = 2\Omega + S_{\rm out} \,,
\end{align} 
where $S_{\text{out}}$ may be holographically given by $S_{\text{bulk}}$ in \eqref{def:S_gen}, it immediately follows from our definitions \eqref{def:2D:class:exp} and \eqref{def:2DQexp} that  
\begin{align}
 \Theta = \theta + \dfrac{1}{2\Omega} \dfrac{d S_{\text{out}} }{d\lambda} \,.
\end{align}
As considered in \cite{Bousso:2015mna}, the QNEC can be derived for expansion-free null surface. When $\theta = 0$, 
eq.~\eqref{focusing-RST} reduces to  
\begin{equation}
 \frac{d\theta}{d \lambda} = - \frac{e^{-4\rho}}{2\Omega} T_{++} \ , 
\end{equation}
where $T_{++}$ is the energy-momentum tensor in the RST model \eqref{QEM}. 
Then, our QFC implies, for $\theta=0$, 
\begin{equation}
 e^{-4\rho} T_{++} \geq \dfrac{d^2S_{\text{out}} }{d\lambda^2} \,.
\end{equation}
This corresponds to the QNEC in two dimensions. In higher dimensions, the QNEC fails for some cases of interest~\cite{Fu:2017lps,Ishibashi:2018ern}. It would be interesting to clarify whether (under what circumstances) the QNEC may possibly fail in two dimensions.

\bigskip
\goodbreak
\centerline{\bf Acknowledgments}

\medskip 
\noindent
This work was supported in part by JSPS KAKENHI Grant No.~JP20K03930, JP20K03938 and also supported by MEXT KAKENHI Grant-in-Aid for Transformative Research Areas A Extreme Universe No.~JP21H05182, JP21H05186. 

\begin{appendix}

\section{Quantum expansion in two dimensions}

We should first note that in two dimensions, there is a subtle issue in the definition of the quantum expansion, as a given co-dimension two surface---whose area $\cal A$ appears in the denominator of the original definition~\eqref{def:Qexp}---now corresponds to a single point in the two dimensional setting. 
This ambiguity is related to the facts that null geodesics can form no co-dimension two surface in two dimensions 
and that the notion of focusing in two dimension is unclear. 
A candidate of the quantum expansion might be simply \cite{Almheiri:2019yqk}
\begin{equation}
 \Theta = \frac{d S_\text{gen}}{d \lambda} \,.
 \label{false}
\end{equation}
In the case of the JT gravity, the QFC for this definition of quantum expansion was shown in \cite{Almheiri:2019yqk}.
However, in our context, 
the gravity part is different from the JT gravity, and hence, 
the definition \eqref{false} does not give desired properties of quantum focusing even in the classical limit.
The counterpart of the classical expansion for \eqref{false} is simply given by 
\begin{equation}
 \theta = \frac{d}{d \lambda} e^{-2\phi} \,.
\end{equation}
As the derivative with respect to the affine parameter 
can be rewritten as ${d}/{d \lambda} = e^{-2\rho}\partial_+$, 
the focusing condition is expressed in the gauge $\rho=\phi$ as 
\begin{equation}
 \frac{d\theta}{d \lambda} = e^{-2\phi}\left(\partial_+ e^{-2\phi}\right)^2 - \frac{1}{2} e^{-4\phi} T_{++} \,,
 \label{false-focusing}
\end{equation}
where we used \eqref{QEM} in the classical limit $c\to 0$. 
Thus, the definition \eqref{false} gives an increasing expansion even for $T_{++}=0$, 
and hence, is not a desirable definition of the quantum expansion.

By inspection, we find it appropriate to replace the geometric area ${\cal A}$ in \eqref{def:Qexp} with the variable $\Omega= e^{-2\phi}+c \phi/24$ introduced in \eqref{def:Omega}. In fact, we have already taken the perspective that $\Omega$ can be viewed as the ``area'' or geometric entropy with quantum correction to define the generalized entropy \eqref{def:S_gen}--\eqref{area}. Another piece of evidence for the correspondence between $\Omega$ in two-dimensions and the area entropy in higher-dimensions can be found in the derivation of two-dimensional black holes in the ``large-$D$ expansion'' in gravity~\cite{Emparan:2013xia,Soda:1993xc}, where the area of the $(D-2)$-sphere in spherically symmetric spacetimes 
is identified with $e^{-2\phi}$ up to a constant factor. In view of this,  we define the two-dimensional counterpart of the classical expansion~\eqref{def:class:exp} as
\begin{align}
 \label{def:CGHS:exp}
 \theta:= e^{2\phi} \dfrac{d}{d \lambda} e^{-2\phi} \,, 
\end{align}
which gives an analogue of the Raychaudhuri equation: 
\begin{equation}
 \frac{d \theta}{d \lambda} = - \frac{1}{2} e^{-2\phi} T_{++} \,.
 \label{focusing-CGHS}
\end{equation}
Thus, the classical focusing theorem holds if the NEC is satisfied. 
The classical expansion \eqref{def:CGHS:exp} would naturally be generalized to the RST model case as 
\begin{align}
 \label{def:2D:class:exp}
 \theta:= \dfrac{1}{\Omega} \dfrac{d \Omega}{d \lambda} \,. 
\end{align}
The derivative of the expansion is expressed in terms of the energy-momentum tensor \eqref{QEM} as 
\begin{equation}
 \frac{d\theta}{d \lambda} 
 = 
 - \left[1- \left(\frac{\Omega}{e^{-2\phi} - {c}/{48}}\right)^2 \right] \theta^2 
 - \frac{e^{-4\phi}}{2 \Omega} T_{++} \ . 
 \label{focusing-RST}
\end{equation}
This corresponds to the Raychaudhuri equation for the classical expansion \eqref{def:2D:class:exp} with quantum correction. In the limit $c\to 0$, eq.~\eqref{focusing-RST} for the RST model reduces to eq.~\eqref{focusing-CGHS} for the CGHS model. 

However, the classical focusing would be violated in the RST model. By using \eqref{field equation RST}, we obtain
\begin{equation}
\label{def:2DRaychaudhuri}
 \dfrac{d \theta}{d\lambda} 
 = 
 - \frac{c}{24} \frac{\left(- \phi - {1}/{2}\right)}
 {\left(e^{-2\phi} - {c}/{48}\right)}  \theta^2
 - \frac{1}{2} \frac{e^{-4\phi}}{\Omega} \left(T_{++}^\text{(cl)} + \frac{c}{12} t_+\right)
 \,, 
\end{equation}
where $T_{++}^\text{(cl)}$ is the classical part of the energy-momentum tensor. We can see from \eqref{def:2DRaychaudhuri} that for two-sided eternal black hole, $t_+=0$, and the classical focusing $ {d \theta}/{d\lambda} \leq 0$ holds as long as the quantum correction in $\Omega$ is sufficiently small, 
namely, for $e^{-2\phi} \gg {c}\phi /{24}$, since 
the first term in \eqref{def:2DRaychaudhuri} is non-positive for $\phi< -{1}/{2}$, 
and non-negative for $\phi>-{1}/{2}$. 
If $c < 48 e$, $\phi> - {1}/{2}$ sufficiently near the boundary $\Omega=\Omega_\text{crit}$. 
Thus, the classical focusing $ {d \theta}/{d\lambda} \leq 0$ holds 
for asymptotic region (large values of $\Omega$), while it could be violated for sufficiently small $\Omega$. 
For the linear dilation vacuum and dynamical black holes, $t_+<0$, we need more careful analysis. 
In any case, the violation of the classical focusing comes from quantum effects. 

\end{appendix}

\end{document}